# Space-time transformation for superluminal signaling

Rui Qi

Institute of Electronics, Chinese Academy of Sciences

17 Zhongguancun Rd., Beijing, China

E-mail: rg@mail.ie.ac.cn

We analyze the possible implication of the existence of superluminal signaling for space-time structure. A new space-time transformation for superluminal signaling is presented based on the superluminal synchrony method. We argue that Lorentz transformation should be replaced by the new transformation in case of the existence of superluminal signaling. Furthermore, we discuss the possible existence of absolute frame, and give a possible practical method to probe it.

## Introduction

As we know, even if relativity theory and present quantum theory permits no existence of superluminal signaling, but its existence has not been refuted in principle yet. It is still unclear whether the complete quantum gravity theory or TOE permits superluminal signaling. In fact, some physicists have guessed that when considering self-consciousness superluminal signaling can be realized in the framework of revised quantum dynamics[1-4]. Especially, some performed experiments may have confirmed the existence of superluminal signaling[5-6].

In this paper, we will mainly analyze the possible implication of the existence of superluminal signaling for space-time structure. A new space-time transformation for superluminal signaling is presented based on the superluminal synchrony method. We argue that the choice of synchrony is no longer a convention, and the superluminal synchrony will be the last choice. This also indicates that Lorentz transformation should be replaced by the new transformation in case of the existence of superluminal signaling. We further discuss the possible existence of absolute frame, and present a possible practical method to probe it.

## Superluminal synchrony method

In this section, we will demonstrate that if there exists superluminal signaling, then it can be naturally used to synchronize the clocks in different positions. We call this method superluminal synchrony method.

As we know, the comparison of different signal speed is irrelevant to the synchrony method of the clocks in different positions. Thus if superluminal signaling or instantaneous signaling does exist, it can be compared with the light propagation in principle. For example, we put two same clocks in different positions A and B. They are not synchronized in the experiment. Then we measure the time interval of different signals propagating between these positions. It can be seen that the larger the propagation speed of the signal the shorter the time interval, especially, the time interval of the superluminal signal will be shorter than that of light signal.

Furthermore, the propagation speed is closely related to the synchrony of clocks in different positions. If we disregard the propagation delay, and directly use the propagated signal to synchronize the clocks in different positions, then the larger signal speed corresponds to more accurate synchrony. Since we can't know the real signal speed before the clocks are synchronized, this kind of inaccuracy is inevitable. Fortunately, if there exists superluminal signaling or instantaneous signaling, then the propagation delay will be zero. Thus we can synchronize the clocks in different positions without error using such superluminal signaling. From a logical point of views, this may be the natural selection of Nature, since superluminal signaling itself implies simultaneity.

Thus we conclude that if there exists superluminal signaling, then it can be used a natural method to synchronize the clocks in different positions, and the simultaneity can be uniquely defined using such superluminal signal. Furthermore, once we synchronize the clocks in all positions, then the propagation speed of other signals can be physically measured. Especially, the light speed or the one-way speed of light can be also measured.

## Superluminal space-time transformation

In the following, we will work out the space-time transformation for superluminal signaling based on the above superluminal synchrony method.

As we know, what has been confirmed in experiments is not the invariance of one-way light speed, but the invariance of two-way light speed[7]. Thus we use the general Edwards transformation[8], which satisfies the invariance principle of two-way light speed, and is irrelevant to the one-way light speed. In frame $S$ we let $c_x = \frac{c}{1-k}$ and $c_{-x} = \frac{c}{1+k}$, which denotes the one-way light speed along x and –x direction respectively, in frame $S'$ we let $c_{x'} = \frac{c}{1-k'}$ and $c_{-x'} = \frac{c}{1+k'}$, which denotes the one-way light speed along x' and –x' direction respectively, where c is the two-way average light speed, $k, k'$ satisfy $-1 \leq k, k' \leq 1$, which represents the directionality of one-way light speed in these two frames respectively. When $k, k' = 0$, the one-way light speed will be isotropy in these two frames. Suppose v is the velocity of S' relative to S along x direction, then the Edwards transformation can be written as follows:

$$x' = h(x - vt) \quad \text{--- (1)}$$

$$y' = y \quad \text{--- (2)}$$

$$z' = z \quad \text{--- (3)}$$

$$t' = h[1 + b(k + k')]t + h[b(k^2 - 1) + k - k']x/c \quad \text{--- (4)}$$

where $h = 1/\sqrt{(1+bk)^2 - b^2}$, $b = v/c$.

Now if there exists superluminal signaling, then we can use the superluminal synchrony method to synchronize the clocks in every frame. Considering the principle of relativity, the instantaneous signal in one frame will be also instantaneous in another frame. This requires that $t' = t$ in the above space-time transformation. Then we have: $[b(k^2 - 1) + k - k'] = 0$; Besides, the superluminal synchrony method will result in that only in one frame the one-way light speed is isotropy, for simplicity we let the frame S to be this isotropy frame, namely we let $k = 0$. Then we get: $k' = b(k^2 - 1) + k = -b$. Thus the above space-time transformation turns to be:

$$x' = \frac{1}{\sqrt{1 - b^2}} (x - vt) \quad \text{--- (5)}$$

$$y' = y \quad \text{--- (6)}$$

$$z' = z \quad \text{--- (7)}$$

$$t' = \sqrt{1 - b^2} \cdot t \quad \text{--- (8)}$$

We call this new space-time transformation for superluminal signaling superluminal space-time transformation.

It can be seen that the superluminal space-time transformation will also result in time dilation and length contraction. The difference with special relativity lies in that the time dilation and length contraction is absolute here, they only refer to the absolute frame S. Besides, the one-way light speed is not isotropy in most frames. In frame with velocity v relative to the absolute frame, the one-way light speed is respectively $c_x = \frac{c}{1 - b}$ and $c_{-x} = \frac{c}{1 + b}$.

It should be denoted that the relativity of simultaneity is only the result of standard Einstein synchrony method, which assumes the isotropy of one-way light speed. The invariance of two-way light speed, which has been confirmed in experiments, is irrelevant to the relativity of simultaneity. What's more, if superluminal signaling does exist, then simultaneity will be absolute, and be irrelevant to the selection of frames.

## The existence of absolute frame

Now we have two kinds of space-time transformations, one is the usual Lorentz transformation in special relativity, the other is the superluminal space-time transformation here. Then which is the real space-time transformation? What is the difference of predictions of these two transformations?

As we see, the key lies in the existence of superluminal signaling. If there doesn't exist superluminal signaling in principle, then there is no difference of predictions for these two transformations. Their different forms result only from the selection of synchrony convention. The Lorentz transformation uses the Einstein synchrony convention, which assumes the isotropy of one-way light speed, and the superluminal space-time transformation uses the superluminal synchrony method, which assumes the absoluteness of simultaneity. It should be denoted that even if superluminal signaling doesn't exist, we can still use other methods to synchronize the clocks in different frames in order to hold the absoluteness of simultaneity, for example, we first synchronize the clocks in one frame using Einstein synchrony convention, then we let the clocks in other frames directly regulated by the clocks in this frame[9]. In fact, we can reach the above conclusion from Edwards transformation more easily. Since these two transformations are just two special forms of Edwards transformation, they will give the same physical predictions in case of the nonexistence of superluminal signaling.

On the other hand, if superluminal signaling does exist, then the above two transformations will represent different physical contents, and give different predictions. The key prediction is the existence of absolute frame. The Lorentz transformation doesn't predict the existence of absolute frame, while the superluminal transformation does predict the existence of absolute frame, in which the measured one-way light speed is isotropy. Besides, the superluminal transformation provides a wider framework than Lorentz transformation. It can not only be used to describe the subliminal and light phenomena, but also be naturally used to describe the superluminal phenomena.

## How to probe the absolute frame?

Now one big problem is still left, i.e. if absolute frame does exist, then how to find it? The direct method is to use the superluminal signaling[1-4]. We can first synchronize the clocks in all frames, then measure the one-way light speed. If we find that the one-way light speed is isotropy in one frame, then we will find the absolute frame, which is just this frame. But even if superluminal signaling can be realized, it is undoubtedly very difficult to obtain it using present technology. Then is there more practical method to find the absolute frame? For this we need to analyze the possible origin of the absolute frame.

The superluminal transformation is deduced based on two assumptions, one is the existence of superluminal signaling, the other is the invariance principle of two-way light speed, which all satisfy the principle of relativity. But the combination of these assumptions does result in the existence of absolute frame, which evidently violates the principle of relativity. This seems to be a paradox. As we think, since the existence of absolute frame mainly results from the existence of superluminal signaling, the reason should also hide in it. The basis of superluminal signaling is the dynamical collapse of wave function, thus the existence of absolute frame should be essentially related to the dynamical collapse of wave function. This implies that the principle of dynamical collapse of wave function may help to find the absolute frame. Let's further look at the dynamical collapse theories.

In one kind of dynamical collapse theory[3][10-15], the nonrelativistic collapse time formula is $t_c \approx \frac{\hbar E_p}{(\Delta E)^2}$. Considering the relativistic transformation of time and energy, the relativistic collapse time formula in any frame should be $t_c \approx \frac{g \hbar E_p}{(\Delta E)^2}$, where $g = \frac{1}{\sqrt{1 - v^2/c^2}}$. This formula contains a term relating to the velocity of frame, which evidently violates the principle of relativity. We may call the frame to which the velocity is relative absolute frame. Thus in the above dynamical collapse theory, the collapse law can indeed help us find the absolute frame. If the measured collapse time in one frame satisfies the formula $t_c \approx \frac{\hbar E_p}{(\Delta E)^2}$, or we find the collapse time of the same wave function in one frame is shorter than that in the neighboring frames, then we can conclude that this frame is just the absolute frame.

## Conclusions

In this paper, we analyze the possible implication of the existence of superluminal signaling for space-time structure. A new space-time transformation for superluminal signaling is presented based on the superluminal synchrony method. We argue that the superluminal transformation provides a wider framework than Lorentz transformation, which should be replaced by the superluminal transformation in case of the existence of superluminal signaling. We further discuss the possible existence of absolute frame, and present a possible practical method to probe it.

## Acknowledgments

Thanks for helpful discussions with Cao Zhang ( University of Maryland ), Yuan-Zhong Zhang ( Institute of Theoretical Physics, Chinese Academy of Sciences ).

## References

[1] E.J.Squires, Phys. Lett. A 163 (1992) 356-358
[2] Gao Shan. LANL e-print quant-ph/9906116 (1999)
[3] Gao Shan, Quantum Motion and Superluminal Communication (Chinese B&T Publishing House, Beijing, 2000)
[4] Gao Shan. NeuroQuantology, Vol.1, No.1 (2003) 4-9
[5] D.Duane and T. Behrendt. Science, 150 (1965) 367
[6] J.Grinberg-Zylberbaum, D.Dalaflor, L.Attie and A.Goswami, Physics Essays, 7 (1994) 422
[7] Yuan-Zhong Zhang, Special Relativity and Its Experimental Foundations. (World Scientific Publishing Co Pte Ltd, Singapore, 1998)
[8] W. F. Edwards, Am. J. Phys., 31, 482 (1963)
[9] Cao Zhang, Found. Phys, 1987
[10] I.C.Percival, Proc. Roy. Soc. Lond. A, 447, 189-209 (1994)
[11] L.P.Hughston, Proc. Roy. Soc. Lond. A, 452, 953 (1996)
[12] D.I.Fivel, Phys. Lett. A 248, 139-144 (1998)
[13] Gao Shan, Physics Essays, Vol.14, No.1, March 2001 (2001)
[14] S.L.Adler, D.C. Brody, T.A. Brun, L.P. Hughston, J. Phys. A 34, 4797-4809 (2001).
[15] S.L.Adler, T. A. Brun, J. Phys. A 34, 8795-8820 (2001).